\documentclass[useAMS,usenatbib]{mn2e}
\usepackage{lscape}
\usepackage{rotating}
\usepackage{amssymb}
\usepackage{gensymb}
\usepackage{natbib}

\usepackage{graphics,graphicx}
\usepackage{epsfig}

\def\msun{\hbox{M$_\odot$}}

\def\t4{\hbox{t$_{\rm 4}$}}

\def\cm3{\hbox{cm$^{-3}$}}
\input psfig.sty

\title[The embedded lifetimes of young massive clusters in M83]
{Studying the YMC population of M83: how long clusters remain embedded, their interaction with the ISM and implications for GC formation theories}
\author[Hollyhead \& Bastian]{K. Hollyhead$^1$, N. Bastian$^1$, A. Adamo$^2$, E. Silva-Villa$^3$, J. Dale$^4$,$^5$, J. E. Ryon$^6$ \\
\& Z. Gazak$^7$ \\
$^{1}$ Astrophysics Research Institute, Liverpool John Moores University, 146 Brownlow Hill, Liverpool L3 5RF, UK\\
$^{2}$ Department of Astronomy, Oscar Klein Centre, Stockholm University, AlbaNova, Stockholm SE-106 91, Sweden\\
$^{3}$ FACom-Instituto de Fsica-FCEN, Universidad de Antioquia, Calle 70 No. 52-21, Medell\'{i}n, Colombia\\
$^{4}$ Excellence Cluster `Universe', Boltzmannstr. 2, 85748 Garching, Germany\\
$^{5}$ Universit\"{a}ts Sternwarte M\"{u}nchen, Scheinerstr. 1, 81679 M\"{u}nchen, Germany\\
$^{6}$ Department of Astronomy, University of Wisconsin-Madison, 475 North Charter Street, Madison, WI 53706, USA\\
$^{7}$ Institute for Astronomy, University of Hawai'i, 2680 Woodlawn Drive, Honolulu, HI 96822, USA\\
}
\date{Accepted 2015 February 12. Received 2015 February 6; in original form 2014 December 19}
\pagerange{\pageref{firstpage}--\pageref{lastpage}}
\pubyear{2015}
\begin{document}
\maketitle
\label{firstpage}
\begin{abstract}
The study of young massive clusters can provide key information for the formation of globular clusters, as they are often considered analogues. A currently unanswered question in this field is how long these massive clusters remain embedded in their natal gas, with important implications for the formation of multiple populations that have been used to explain phenomena observed in globular clusters. We present an analysis of ages and masses of the young massive cluster population of M83. Through visual inspection of the clusters, and comparison of their SEDs and position in colour-colour space, the clusters are all exposed (no longer embedded) by $<$ 4 Myr, most likely less, indicating that current proposed age spreads within older clusters are unlikely. We also present several methods of constraining the ages of very young massive clusters. This can often be difficult using SED fitting due to a lack of information to disentangle age-extinction degeneracies and possible inaccurate assumptions in the models used for the fitting. The individual morphology of the H$\alpha$ around each cluster has a significant effect on the measured fluxes, which contributes to inaccuracies in the age estimates for clusters younger than 10 Myr using SED fitting. This is due to model uncertainties and aperture effects. Our methods to help constrain ages of young clusters include using the near-infrared and spectral features, such as Wolf-Rayet stars. 

\end{abstract}
\begin{keywords} 
\end{keywords}

\section{Introduction}
\label{sec:intro}

The majority of stars in our universe do not form alone, but rather in groups, or clusters \citep{lada03}. Therefore, it is necessary for us to gain an understanding of how these clusters have formed and evolved. %There are a variety of types of clusters, from very low mass open clusters living in the discs of galaxies to the highest mass, dense globular clusters residing in halos. 
The question addressed in this paper concerns the early phases of cluster evolution, specifically how long the clusters remain embedded. The stars belonging to the cluster will form from a mass of cold gas, but how long after this initial stage does the cluster keep leftover gas from the formation process bound to it?  

To investigate this question we study the young massive cluster population of M83. These young massive clusters, or YMCs, are often considered to be the precursors to globular clusters (GCs) (e.g. \cite{sch98} and \cite{krui15}), once only thought to have formed in the early universe. Considering these objects as analogues, we can study YMC formation to understand how GCs formed and how they evolved to the state we see them today. 

The host galaxy of our clusters, M83, is a nearby quiescent spiral at $4.5$ Mpc \citep{thim03}, with a rich cluster population that has been widely studied, for example, \cite{ba12},  \cite{chand10} and \cite{whit11} are only a few recent papers on this topic. M83 has been widely studied partly due to this rich population of clusters (7290 clusters and possible associations within the catalogue used in this paper from \cite{silva14}), which means in depth studies can be carried out and population statistics are meaningful. There are also clusters of a wide range of masses and ages. Additionally, M83 is thought to be very similar to our own Milky Way galaxy, and so any results regarding the cluster population of M83 can be used to infer results for the Milky Way, whose complete cluster population cannot be determined or studied due to the clusters' and our position in the disc \citep{larsen08}.

Studies of other young massive clusters both within and outside the Milky Way show the importance of feedback within star forming regions and clusters and give indication of the timescales for gas to be expelled from clusters. For example, the Arches cluster towards the centre of our galaxy has an age of $2.5-4$ Myr \citep{blum01} and is observed to be free of gas. Other clusters of similar masses, such as NGC 3603 (aged $\sim1-2$ Myr) or Westerlund 1 (aged $\sim3-6$ Myr) \citep{neg10} \citep{gen11} \citep{brand08} also have no dense gas bound to them. It has been found that even for lower mass systems than those already mentioned, the clusters can potentially remove gas in $<1$ Myr \citep{seale12}. However this is not the case for all lower mass regions, as some seem to require more time to clear gas, such as W3-main (e.g. \cite{bik14}) or NGC 346 in the SMC (e.g. \cite{smith08}). It has also been found that feedback can prompt further star formation around the central source, such as in the Ruby Ring region \citep{adamo12} or Cluster 23 in ESO 338 \citep{ostlin07}.     

This study focuses on the H$\alpha$ morphology around the YMCs. H$\alpha$ emission traces the gas associated with the clusters, so through visual inspection of how the H$\alpha$ is orientated, it is possible to determine whether gas is still contained within the cluster or has been expelled. By comparing the H$\alpha$ morphology with the age of the clusters we can find the age by which the clusters have removed any remaining gas.

This technique has previously been used by \cite{whit11}, who investigated clusters in Field 1 of M83, as shown in Fig.~\ref{fig:fields}. They used H$\alpha$ morphology as a potential age dating technique for clusters, rather than specifically studying the length of the embedded phase. This study provided an excellent starting point for Field 1, which we now expand into the other 6 fields.  

In \S~\ref{sec:obs} we discuss the data used for the study and how this was used to identify and classify clusters. Additionally, we discuss issues with age dating young clusters and how we tried to mitigate this problem. \S~\ref{sec:results} describes our results for the length of the embedded phase for clusters and other methods for constraining cluster ages. In \S~\ref{sec:discussion} we discuss the implications of the results for GC formation scenarios, in particular the Fast Rotating Massive Star scenario (FRMS) while in \S~\ref{sec:conc} we give our conclusions.

\section{Observations and Techniques}
\label{sec:obs}

\subsection{Data}
\label{sec:data}

\begin{figure}
\centering
\includegraphics[width=8.2cm]{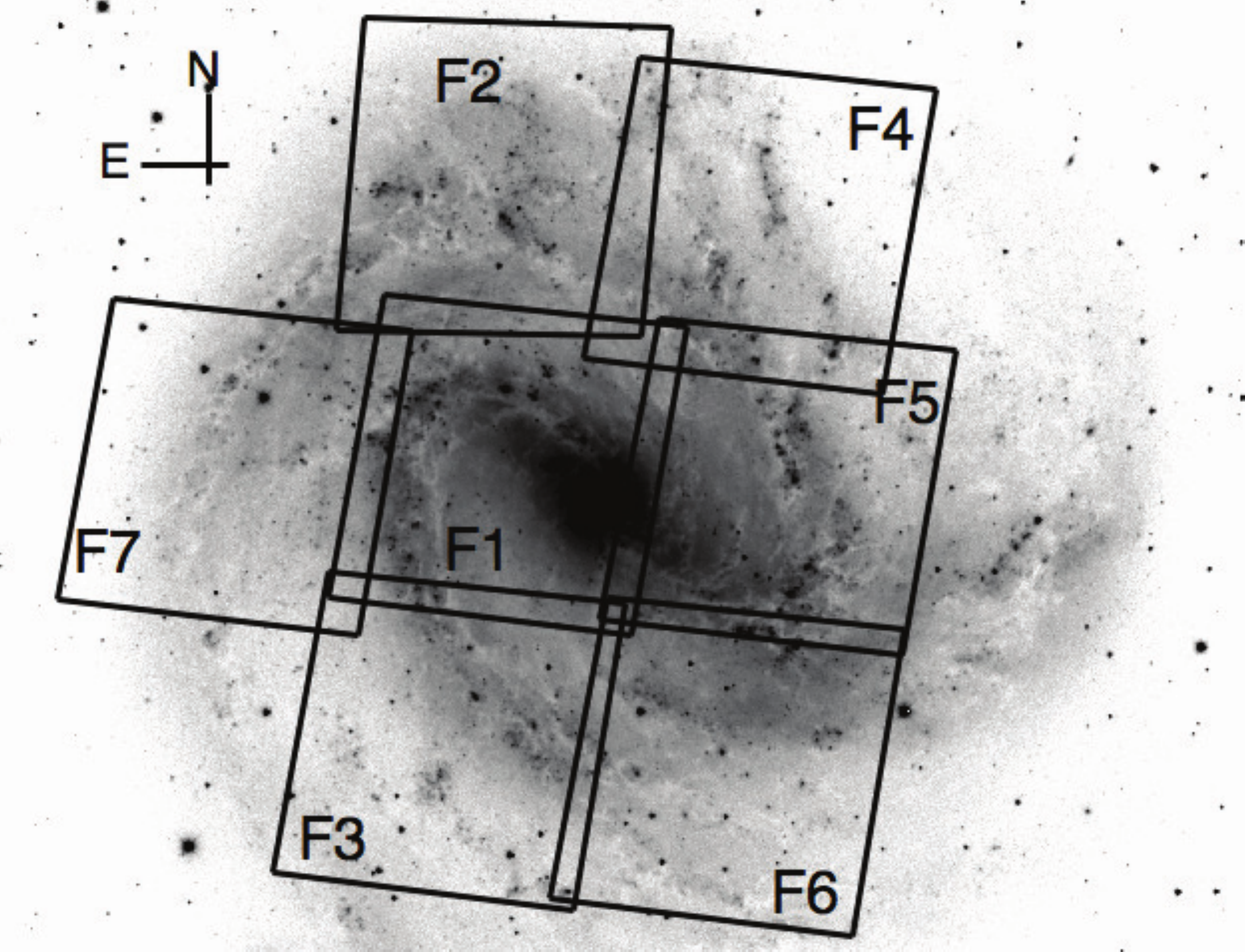}
\caption{The orientation of the seven M83 fields in HST WFC3. As can be seen, the majority of the galaxy is covered, with a few outer regions missed. Image taken from Adamo et al. (in prep).} 
\label{fig:fields}
\end{figure}

Studying the cluster population of M83 requires sufficiently high quality data to resolve individual clusters. In addition, there is a further constraint as the detail of the H$\alpha$ morphology needs to be visible around each cluster. In order to satisfy these requirements, HST WFC3 data was used, as it provides sufficient resolution at the distance of $4.5$ Mpc. The images were obtained from the publicly available HST Legacy Archive for all seven fields of M83 in the U (F336W), B (F438W), V (F547M) (F555W for Fields 1 and 2), H$\alpha$ (F657N), I (F814W) and H (F160W) bands. The archive provides fully reduced and drizzled data. The orientation of the seven fields is shown in Fig.~\ref{fig:fields}, taken from Adamo et al, in prep. No transformation was made in the data to the Cousins-Johnson filter system, however we adopt U, B, V, H$\alpha$, I and H for shorthand, as the WFC3 filters correspond approximately to these wavelengths. 

\cite{ba12} and \cite{silva14} used these images to identify an extensive catalogue of clusters within M83 using a combination of automated and manual procedures. Photometry in the U, B, V, H$\alpha$ and I bands was already available and we added H band magnitudes for this study. Photometry was performed for each of the existing clusters on WFC3 IR channel H band images, and added to the catalogue. The coordinates were transposed to the H band images using the xyxy transform in the astronomy library for IDL, and DAOphot was used to find the magnitudes with a 1.5 pixel aperture and 2.4 pixel background apertures with annuli of 0.9 pixels. The pixel size of the IR detector is 0.13", giving aperture sizes of $\sim0.2$", or $\sim4.2$ pc. We chose this small aperture size so that they were physically the same size as the 5 pixel aperture used for the other bands, ensuring that there would be no effect from spatial differences. Some of the sources were not bright enough in the H band to calculate magnitudes and the H band images covered a slightly smaller area of sky than the other filters, so several clusters outside the limit of the H band had unreliable photometry. Despite these issues, $\sim5600$ of the 7290 clusters in the catalogue were updated with H band magnitudes.

After removing all sources with poor or no photometry, an aperture correction of 0.62 (calculated from a sample of approximately 20 sources) was subtracted to correct for lost flux due to a small aperture. Furthermore, Fields 1 and 2 were dithered differently to the other 5 fields. Our very small apertures were highly susceptible to this difference, resulting in a fairly significant reduction to the magnitudes in Fields 3 to 7. We corrected for this by subtracting 0.6 from the magnitudes in those fields. This value was found from comparing magnitudes of the same sources in the overlapping regions between Fields 2 and 4. The H band photometry is used as a method for age dating clusters, as described in \S~\ref{sec:hband}.

\subsection{Cluster selection}
\label{sec:clusters}

As part of the selection process \cite{ba12} and \cite{silva14} used concentration indices (CIs) to refine the sample. A CI is a measurement of how centrally concentrated emission is for a source. This value helps distinguish between field stars and clusters; stars should be highly centrally concentrated as point sources, whereas clusters have extended emission. Star contaminants can therefore be removed by cutting sources below an index selected from histogram plots. A histogram plot of the concentration index for each source demonstrates two defined sections: the first is a sharp peak consisting of individual stars followed by a lower peak with a smooth tail that illustrates the cluster population. Removing all sources with values in the first peak leaves the cluster candidates. These candidates were then visually inspected to determine if the source was a reliable cluster (class 1), potentially a cluster (class 2) or an incorrectly identified as a cluster in the previous steps (class 3). The last of these populations were removed. Error cuts were also applied to remove sources with poor photometry and aperture corrections were applied to account for flux missed by the aperture. We used the most recent updated version of this catalogue from \cite{silva14}, which included photometry in the U, B, V, H$\alpha$ and I bands as well as estimates of the age, mass and extinction ($A_{V}$) for each cluster. These properties were obtained by comparing each cluster's spectral energy distribution (SED) with models of simple stellar populations as per \cite{a10ab}. 

To identify the possible YMCs that could be considered GC progenitors, age and mass cuts were applied to the catalogue. No distinction was made between class 1 and 2; all were included in our final selection. Our final sample consisted of 91 clusters that were younger than 10 Myr and had a mass $>5000$ \msun. The lower mass cut was applied to minimise the effect of stochastic sampling.  %The majority of these clusters had ages of $\sim6$ Myr and masses of $\sim10^4$ \msun\ up to a maximum of $10^{4.8}$ \msun . Though the typical masses of YMCs discussed in the literature are of the order $\sim10^6$ \msun, it is feasible that the higher mass clusters in my sample could survive disruptive processes to become the $10^4$ \msun\ GCs that we see today.   

\subsection{Cluster categorisation}
\label{sec:bubbles}

\begin{figure}
\centering
\includegraphics[width = 5cm]{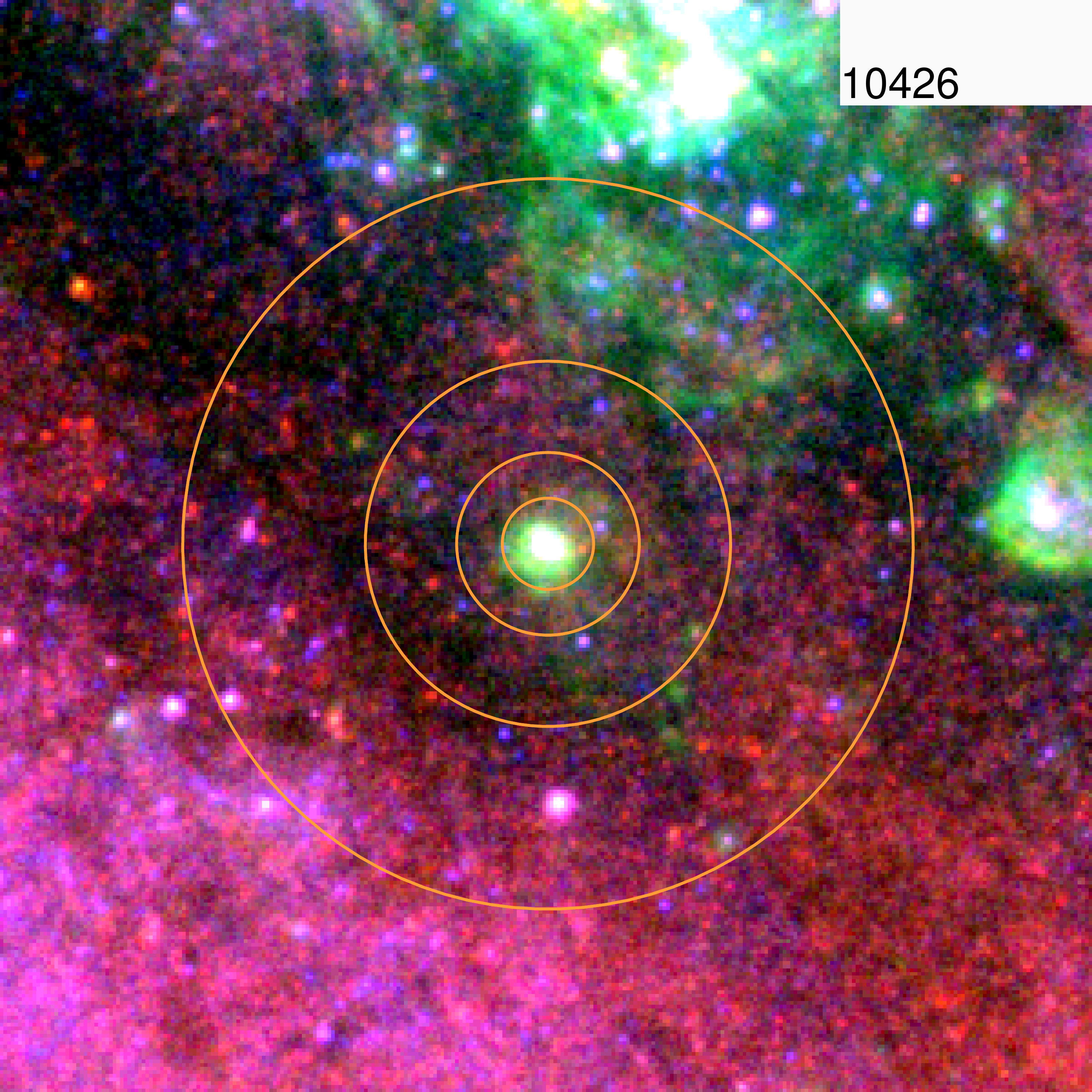}
\includegraphics[width = 5cm]{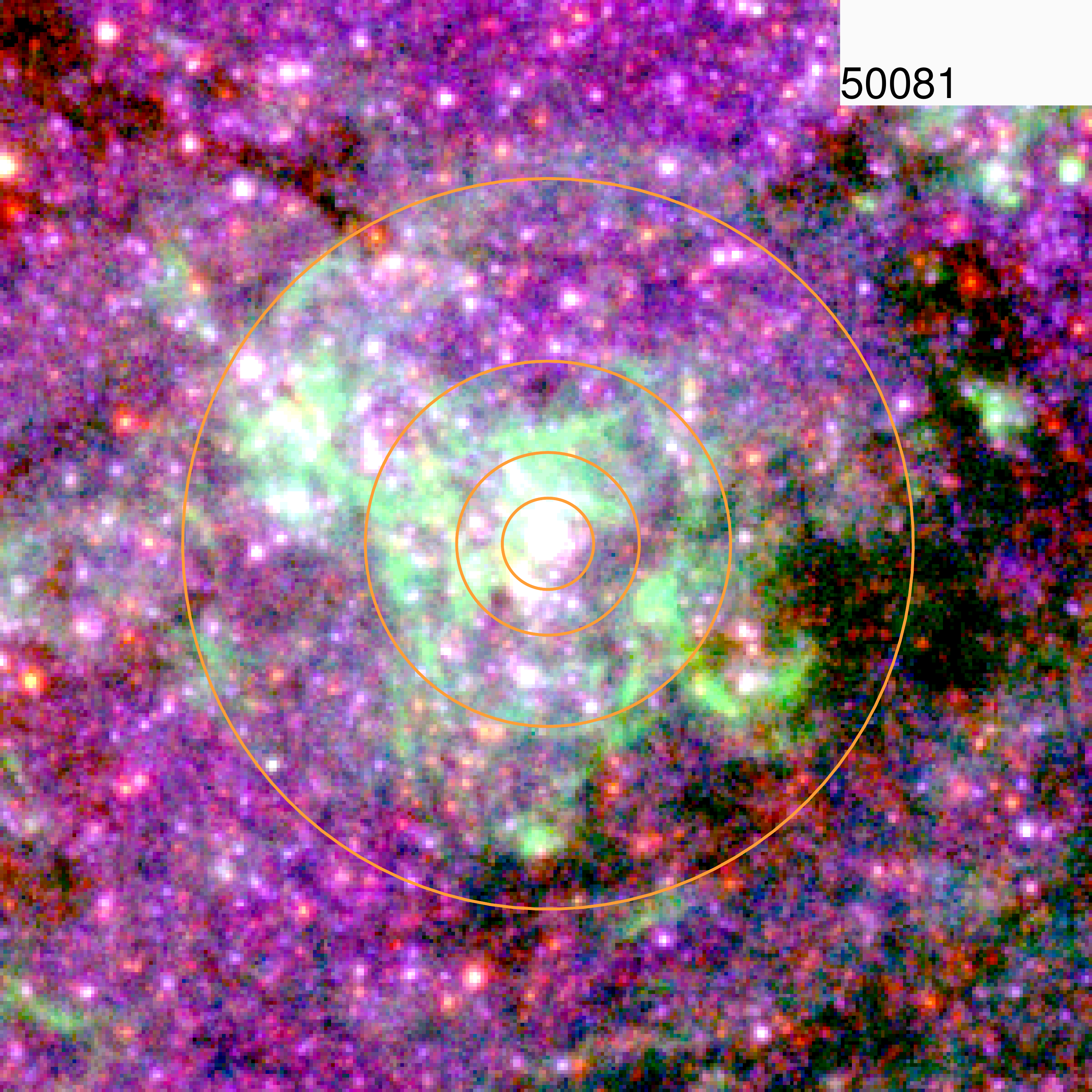} 
\includegraphics[width = 5cm]{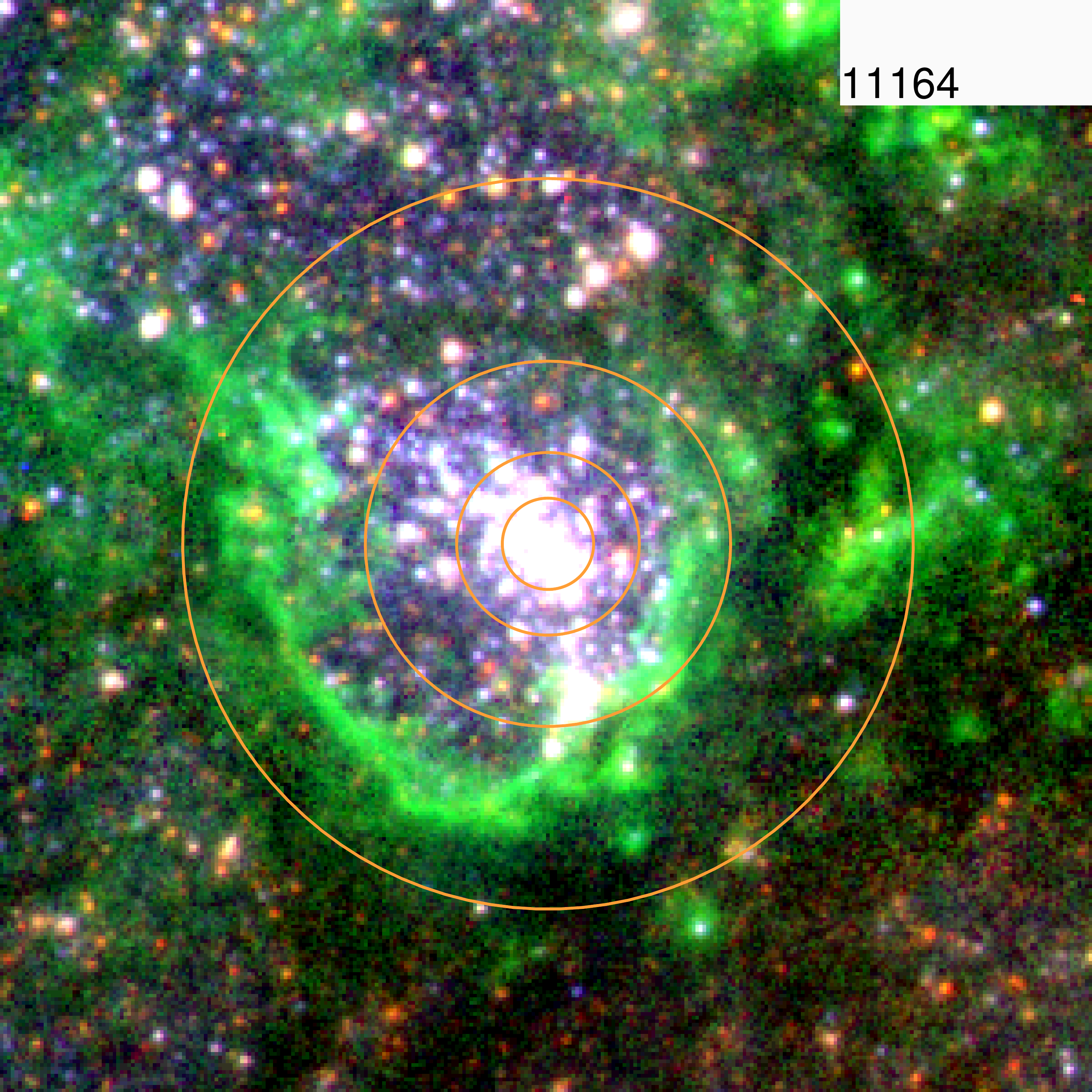}
\caption{Examples of clusters in different categories. The number in the right hand corner gives the cluster's ID within the catalogue and the concentric circles indicate radial distances from the centre of the cluster to 10, 20, 40 and 80 pc respectively, to indicate how far the cluster has cleared gas. The top image is an embedded cluster, the second cluster is partially embedded and the bottom cluster exposed, shown by the clear bubble. Blue colour traces B band emission, green traces H$\alpha$ and red I band.}
\label{fig:clusterexamples}
\end{figure}

After identifying the clusters, we determined whether each would be classed as embedded (no shells or areas where gas had been cleared), partially embedded (part of the cluster had cleared gas) or exposed (clear bubble around the cluster, possibly surrounded by a shell of ionised gas). Categorisation of the clusters was carried out by visual inspection of the H$\alpha$ morphology associated with each cluster. Images of each cluster were created by cropping the original fits files to a 12 x 12 arcsecond area with the cluster at the centre. After creating images in each band, the B, H$\alpha$ and I bands were stacked to create colour images for each cluster, clearly displaying the position of gas in the cluster (traced by the H$\alpha$ emission). H$\alpha$ continuum subtracted images (provided by William Blair, and as discussed in \cite{blair14}) were used to confirm whether the cluster was embedded or not, if the colour image was unclear. 

Examples of each category are shown in Fig.~\ref{fig:clusterexamples}. The first cluster is classed as embedded, as it is clearly surrounded by gas (represented in green), which is reinforced by a high $A_{V}$ of 2.70, caused by dust associated with the gas. SED fitting gives the cluster a very young age of $\sim4$ Myr, and therefore it is reasonable to expect that the cluster is still embedded, although we will return to issues of age fitting in \S~\ref{sec:nohafit}. The second cluster is partially embedded, with a lower $A_{V}$ of 0.39 and an age estimate of $\sim7$ Myr. These intermediate clusters provide a snapshot of the way in which the clusters have removed their gas to progress to an exposed phase. They can be used to provide an estimate of the age at which clusters begin to expel their excess gas and indicate that clusters above this age should be at least partially free of gas. For this reason, they could be considered as the most important clusters in the sample, however the classification between partially embedded and either embedded or exposed is sometimes difficult to discern, depending on how advanced the clearing is for the cluster. A cluster almost entirely devoid of gas may appear to be only slightly embedded but the gas may not be associated with the cluster at all. The final cluster is exposed with an age of $\sim6$ Myr, as shown clearly by the bubble and shell of ionised gas, which has been cleared to a distance of $> 80$ pc on the northern side, but only $\sim30$ pc on the southern side. This is an example of a `blowout'.

The poor quality of the photometry (offset location on the colour-colour diagram possibly caused by contamination from neighbouring sources) for some clusters produced wrong estimates for their physical properties via SED fitting. These outlier sources could also be dominated by stochastic IMF sampling, which is known to produce extreme offsets with respect to traditional stellar evolutionary tracks \citep{fous10}. Therefore we excluded these sources.
 %Several exposed clusters lay in areas which infeasible compared to the model track, leading to the closest incorrect age and higher mass to be assigned to the cluster, along with extremely high $A_{V}$ values.
After identification they were confirmed to be unreliable sources from visual inspection of the images, where all the clusters appeared to have small radii and were red in colour. This may have been due to contamination from nearby sources. Several further clusters were removed due to poor quality images, possibly caused by the continuum that is also detected in the H$\alpha$ band, meaning that the orientation of the H$\alpha$ could not be accurately determined from visual inspection. After these clusters were removed from the subsample, the final confirmed YMCs numbered at 35 exposed, 16 partially embedded and 15 embedded clusters.  

\begin{figure}
\includegraphics[width=8.5cm]{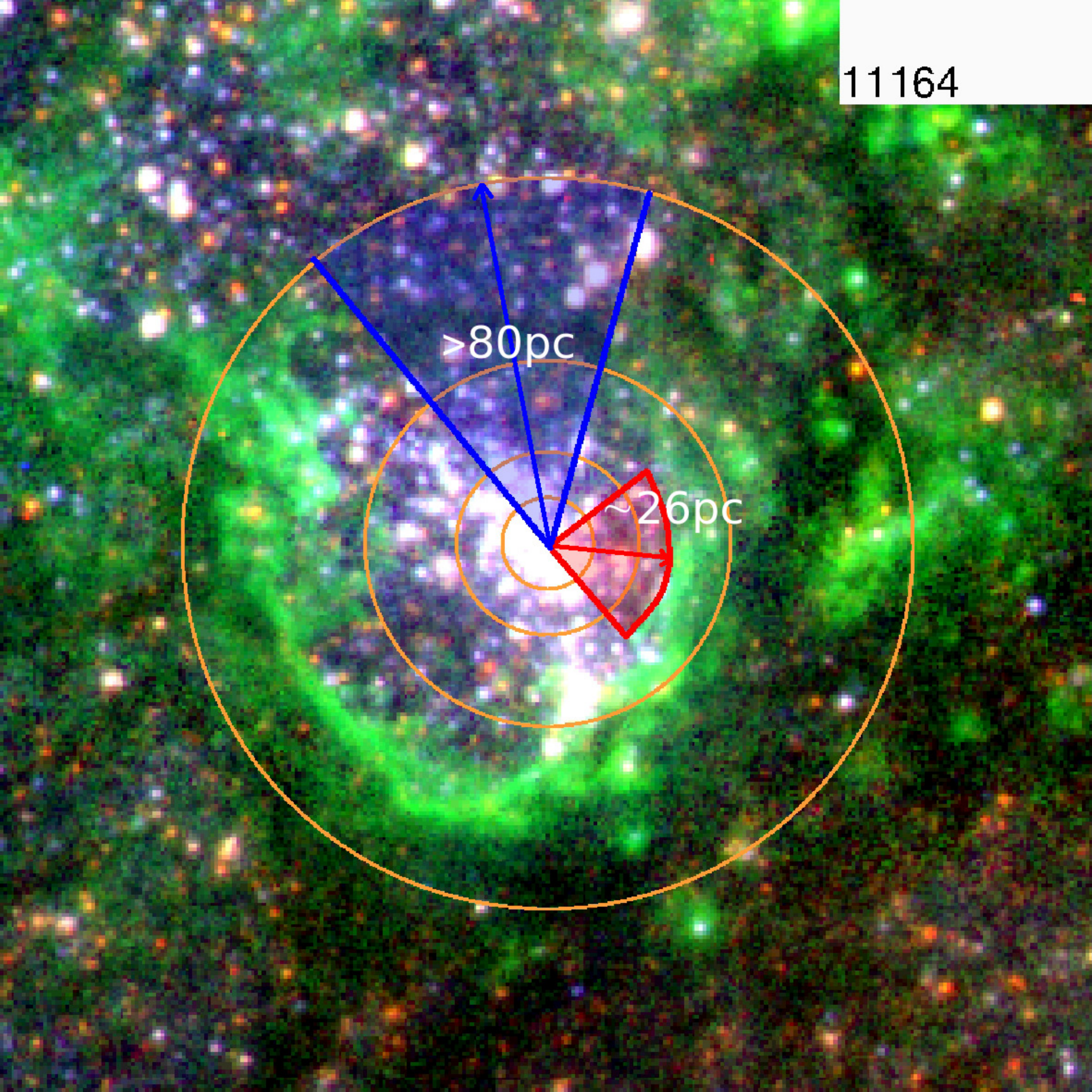}
\centering
\caption{Illustration of the process used to obtain an indication of the extent of clearing of the gas from the clusters. The two shaded areas show the angles that were measured for the minimum and maximum distances cleared in red and blue respectively, labelled with the distance. Angles and pixel lengths were measured in GNU Gimp.}
\label{fig:measureexp}
\end{figure}

After identifying the state of embedded-ness for each cluster, each of the partial and exposed clusters were studied again to find the extent of the gas cleared. Using measurement tools in the GNU Gimp image processing program, the distance from the centre of the cluster to the closest point on the inner bubble edge was measured in pixels. The distance to the part of the inner bubble edge furthest from the centre was measured in the same way, unless the distance exceeded the largest circle on the image, in which case the distance was recorded as $>$ 80 pc, for simplicity. Pixel measurements were converted to parsecs to give the distances. Finally, as an indication of the extent to which the bubbles had been cleared, the angular area of the circular region surrounding the cluster that had been cleared to these maximum and minimum distances were measured using tools in GNU Gimp, as shown in Fig.~\ref{fig:measureexp}. If there were multiple regions cleared to similar distances, the angles measured for each were summed. These angular areas were converted into percentages of the area surrounding the cluster by dividing by 360$\degree$, as included in Table~\ref{table:sample}. This information was used to compare with masses and determine whether a correlation exists between mass of the cluster and the size of the bubbles, as discussed in \S~\ref{sec:blowouts}. Additionally, the percentage information could potentially indicate the existence of a blowout in the gas. We did not measure inner and outer shell radii for embedded clusters. 

Two of the partially embedded clusters in Table~\ref{table:sample}, as indicated by $^\ast$, actually have no maximum distance cleared. These two clusters initially look embedded and are surrounded by gas, however, after inspection of the clusters' SEDs and their position on the colour-colour plot (as discussed later in \S~\ref{sec:nohafit} and \S~\ref{sec:ccplot} respectively) it was clear that they were not completely obscured by gas, and had likely cleared an area in our line of sight. This caused them to appear bluer and could be accurately identified as partially embedded. 

\subsection{Refined age estimates of young clusters}
\label{sec:nohafit}

\begin{figure}
\includegraphics[width = 8.5cm]{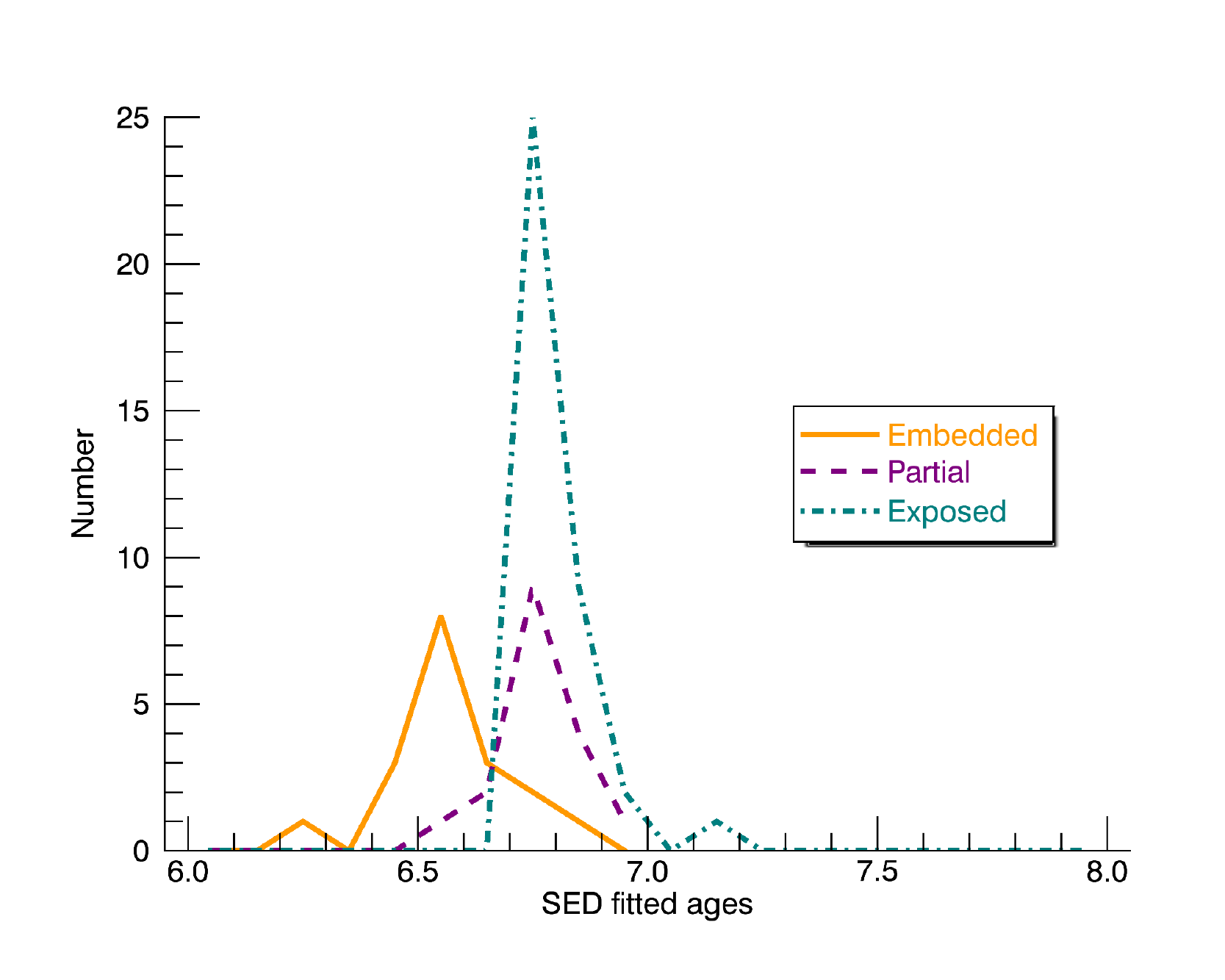}
\caption{The distribution of ages for the young massive clusters using SED fitting of U, B, V, H$\alpha$ and I bands. The orange solid line is embedded clusters, the purple dashed line is partially embedded clusters and the teal dot-dashed line are exposed clusters.}
\label{fig:agedist}
\end{figure}

Fig.~\ref{fig:agedist} shows the distributions of SED fitted ages for embedded, partially embedded and exposed clusters. It is important to note that the SSP models used to fit the ages include nebular emission, which should improve accuracy, and an assumed covering fraction of 50\% (50\% of the ionising photons are absorbed within the aperture). Additionally these ages can be affected by an age/extinction degeneracy and model uncertainties due to factors such as covering fraction or metallicity, for example. As demonstrated, the embedded population peaks at a younger age than the partially embedded and exposed clusters, as we would expect. The partially embedded distribution peaks at a more advanced age than would be expected, at the same age as the exposed population. This could indicate that the process of removing gas is very rapid, or just that the earlier stage partially embedded clusters appear embedded and have been wrongly identified. Alternatively, this could be an effect of binning the ages into the distribution. As this effect is H$\alpha$ morphology dependent, we found that this could be due to an issue with H$\alpha$. 

The clusters in this sample have been age dated using SED fitting of U, B, V, H$\alpha$ and I bands. The degeneracies associated with this method can give inaccurate estimates for the ages of young clusters less than $\sim10$ Myr old. While using all five of the aforementioned bands for the fitting provides the best age estimates for the majority of the clusters, including H$\alpha$ presents a problem with some of the partially embedded and exposed clusters. 

\subsubsection{Colour-colour plot}
\label{sec:ccplot} 

\begin{figure}
\includegraphics[width = 8.5cm]{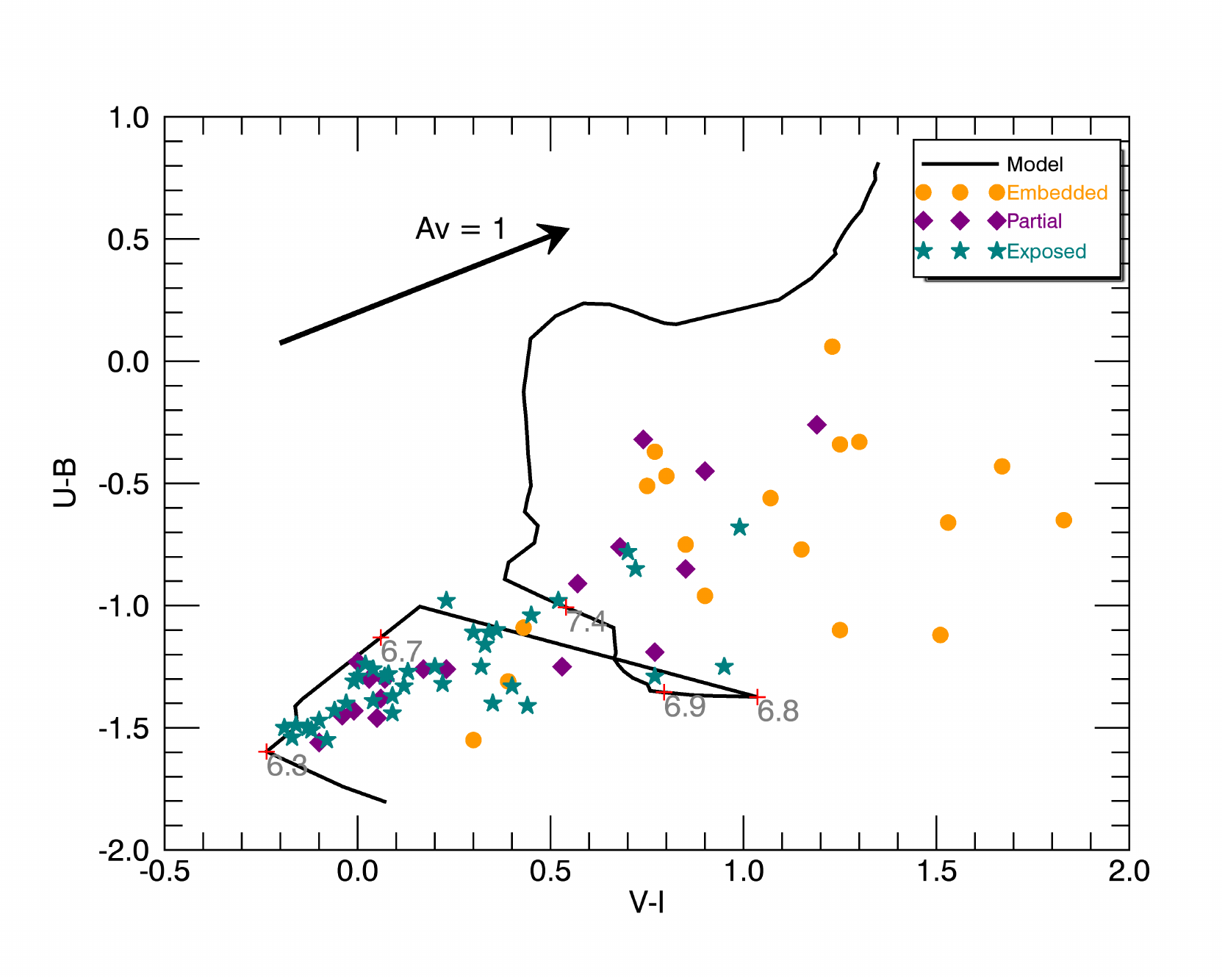}
\caption{Colour-colour plot of U-B vs V-I for the final sample of YMCs in M83. The model track by Zackrisson et al, 2011 is overplotted to indicate where on the evolutionary track the clusters would lie. Orange circles indicate embedded clusters, purple diamonds are partially embedded and green stars are exposed clusters.}
\label{fig:ccplot}
\end{figure}

An important tool in determining the ages of clusters, which can also be used to improve age estimates and that displays the problems experienced in using SED fitting is the colour-colour plot. Fig.~\ref{fig:ccplot} shows this plot for the YMC sample of M83. The model plotted over the points is the cluster evolutionary model for M83 at solar metallicity that includes nebular and continuum emission and uses a Kroupa IMF from \cite{zack11}, which was also used for SED fitting of the clusters. The orange circles correspond to embedded clusters, the purple diamonds to partials and green stars to exposed clusters. Additionally, we have labelled several of the logarithmic ages along the model track. There are some key differences between these populations, which are clearly visible on the plot, and which can provide further information about the early evolution of these clusters. 

The embedded clusters lie primarily in the top right hand corner of the distribution. This is expected, as they all have young ages so should lie on the lower end of the model, but are highly extincted due to their surrounding gas and dust and are moved to the top right section. The exposed clusters also lie where we would expect, in the bottom left of the diagram. This indicates that they are still very young, however, unlike the embedded clusters, a lack of gas means they have much lower extinction and so are much closer to their position on the model. The partially embedded clusters lie scattered around the plot, with extinction varying due to the orientation of the exposed part of the cluster. If the cluster has cleared our line of sight from gas, then the extinction will be lower than a cluster that has cleared in another direction. The position of points on the graph shows a positive result: the classifications of clusters using the colour-colour plot agree with those from the morphological analysis. 

The colour-colour plot can also provide information about the age that clusters start to remove gas. In the region dominated by the exposed clusters, they appear to form a line which is approximately parallel to the extinction vector. This line could be showing the change in the amount of gas cleared over time. If the line is followed back along the extinction vector to the model track, it indicates that the clusters initially began to remove gas at an age of $\sim2$ Myr, depending on the model. 

This plot also indicates why SED fitting can experience issues when fitting ages for embedded and partially embedded clusters. The clusters are brought back along the extinction vector until they meet the model, giving estimates of the age and amount of extinction for the cluster. The shape of the model track introduces a potentially problematic degeneracy for embedded and partially embedded clusters in the top right of the distribution. 

The plot indicates that clusters begin to expel gas at a few Myr, and it only takes this long to create a shell HII region. The embedded clusters have compact HII regions, indicating that they should be very young, and therefore be fitted ages on the earliest section of the model track, however they can be fitted to three different sections of the track due to the degeneracy, assigning them lower extinctions and older ages than they actually have. Partially embedded clusters can also be susceptible to this problem. This age-extinction degeneracy can only be resolved with accurate knowledge of the UV slope \citep{cal14}. 

\subsubsection{The problem with H$\alpha$}
\label{sec:ha}

\begin{figure}
\includegraphics[width = 8.5cm]{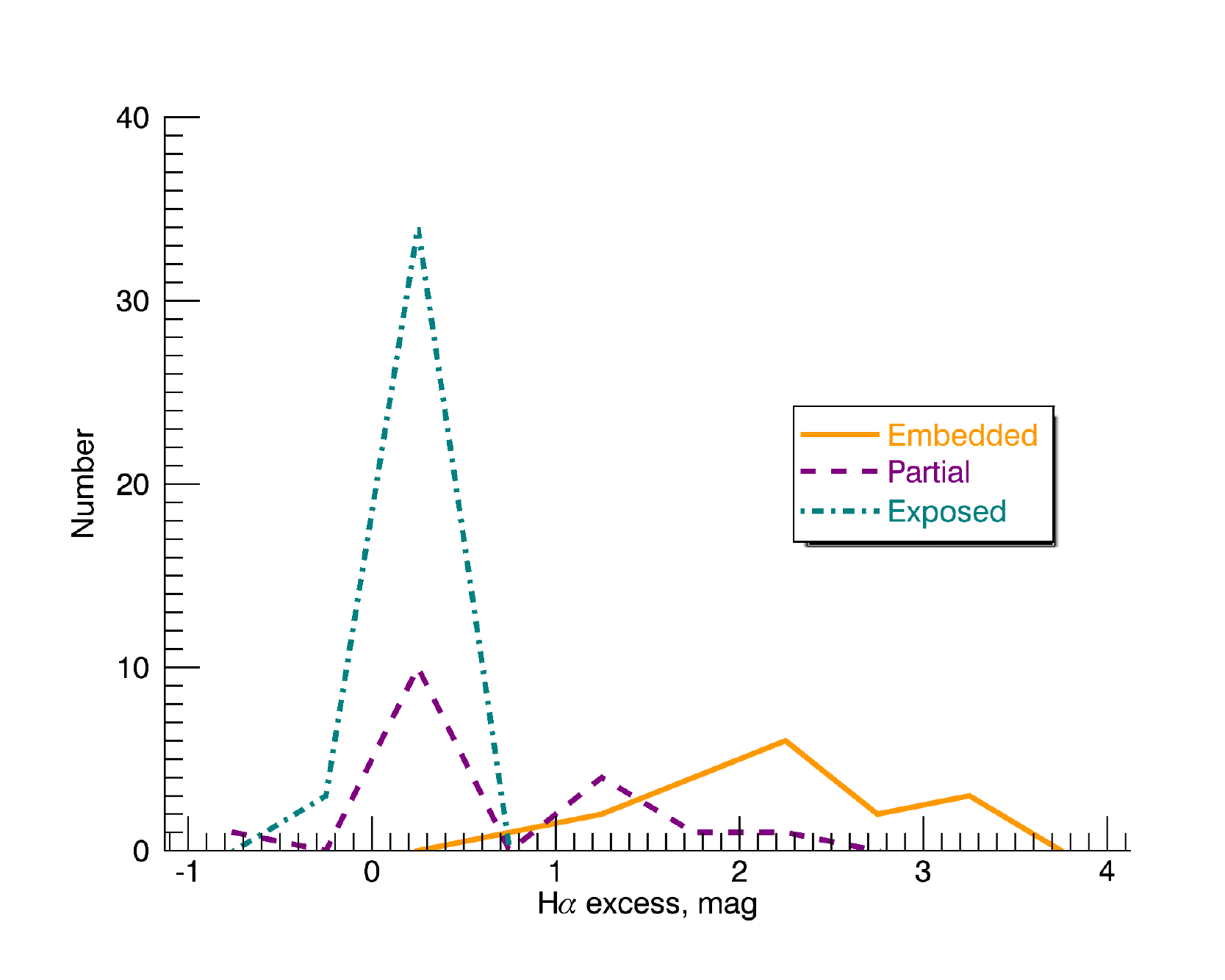}
\caption{A histogram plot of H$\alpha$ excess for embedded, partially embedded and exposed YMCs in the sample. The orange solid line is embedded clusters, the purple dashed line partially embedded clusters and the teal dot-dashed line is exposed clusters. They each peak at different levels of excess, with partially embedded clusters split into 2 peaks, indicating differences in the H$\alpha$ morphology.}
\label{fig:hist} 
\end{figure}

Fig.~\ref{fig:hist} shows an approximation of the H$\alpha$ excess for the three categories of clusters. The H$\alpha$ excess is the average of the magnitudes in V and I minus H$\alpha$ for the cluster. This value provides an indication of the how much the emission from the gas has contributed to the H$\alpha$ magnitude, by subtracting the contribution from the stars (approximated by $(V+I)/2$, as no other bands nearer H$\alpha$ were available). Using $(V+I)/2$ accounts for extinction and offers a close alternative to the actual continuum.

When using SED fitting to approximate ages, young clusters would be expected to have a certain amount of H$\alpha$ emission as they may still contain or be surrounded by ionised gas. As the figure shows, the exposed clusters peak at close to zero excess. This means that even for newly exposed clusters with the expelled gas still nearby, no contribution is made from the gas to the H$\alpha$ magnitude. This would suggest that these clusters have cleared gas to a radius larger than the aperture used for the photometry at a young age. By using H$\alpha$ in the fit for the age, the cluster appears older due to a lack of flux in this band, despite still being young enough to produce radiation that ionises the surrounding gas. In this case, the gas is arranged in a bubble surrounding the cluster, and therefore the majority of the H$\alpha$ emission is from a shell outside of the aperture. If the clusters are exposed but there is still H$\alpha$ emission within the aperture from the nearby gas, a fit using H$\alpha$ may still be more accurate than without. \cite{ba14ab} found a similar effect in the nebular continuum in the near-infrared when studying YMCs.

The partially embedded clusters have 2 peaks in the histogram in Fig.~\ref{fig:hist}. This indicates 2 populations: those that have cleared gas in our line of sight and have close to zero H$\alpha$ excess, and those that still have contribution from the gas. Each population would require different fits, sometimes including H$\alpha$ and others not. 

The embedded clusters all have H$\alpha$ excess as the gas is still within the cluster, or closely nearby and therefore is always included in the measurement of the magnitude. It is unlikely that any embedded clusters would require a fit without H$\alpha$. 

One of the main issues with H$\alpha$ is the small apertures that are used to obtain photometry. They are ideal for capturing the continuum emission, but not the entire emission from the region, so the H$\alpha$ fluxes can be underestimated. This loss of flux cannot be accounted for in the aperture correction, as the morphology of the gas surrounding the clusters that may contribute to the H$\alpha$ flux varies widely. This also means that the excesses displayed in Fig.~\ref{fig:hist} are heavily dependent on the size of the aperture used. Contamination from neighbouring H$\alpha$ sources can also affect the flux in crowded regions. 

%The exposed clusters with little or no excess correspond to where the continuum dominates. The embedded clusters show excess and H$\alpha$ emission, though we cannot quantify how much flux is lost from the small aperture used.

\subsubsection{Accounting for the problem}
\label{sec:fixha}

\begin{figure}
\includegraphics[width = 8.5cm]{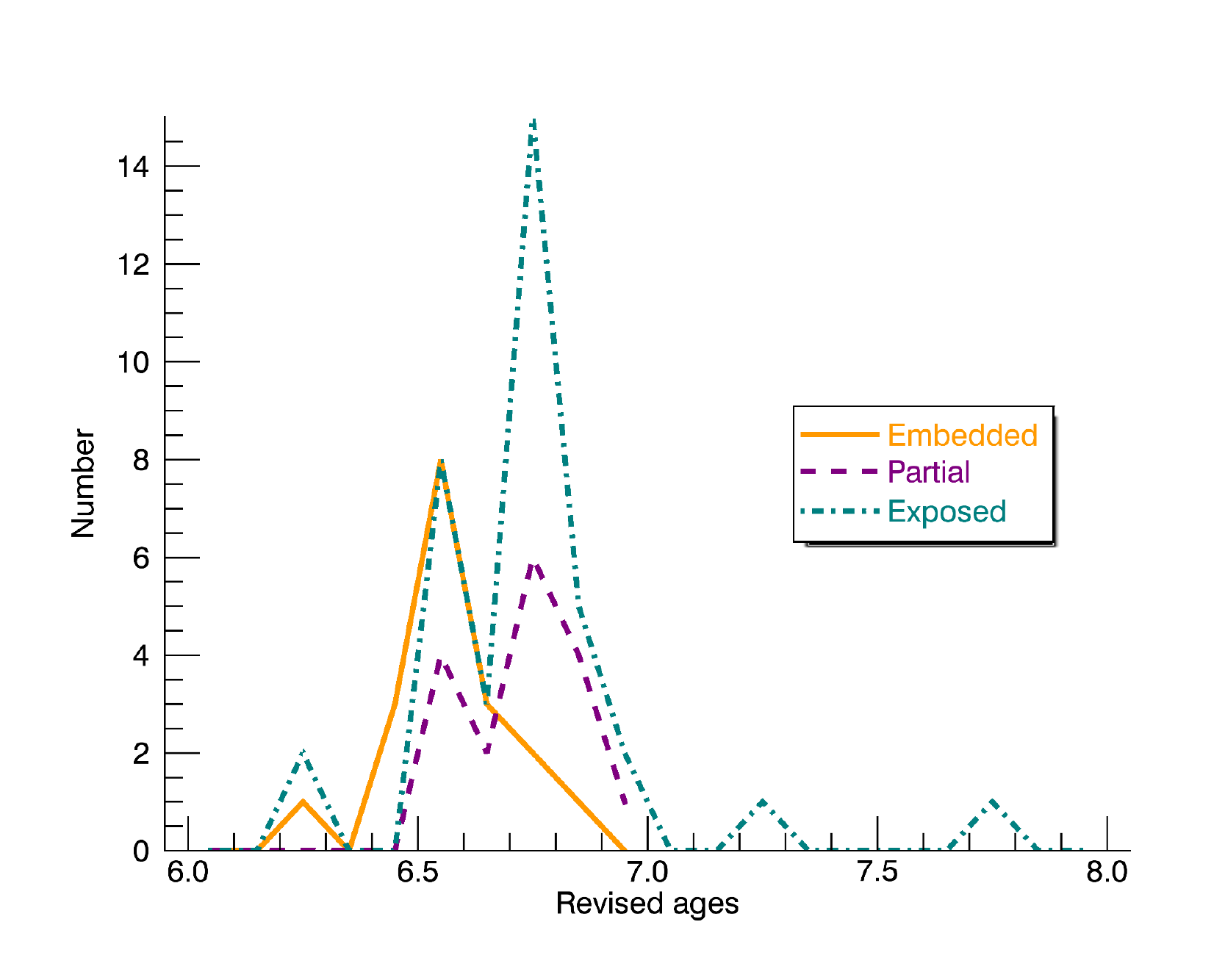}
\caption{New age distribution for the cluster sample, after selecting the correct fit from each cluster's SED, position on the colour-colour plot and H$\alpha$ morphology. As before, the solid orange line is the embedded population, the dashed purple line represent the partially embedded sample and the teal dot-dashed line is the open clusters.}
\label{fig:newages}
\end{figure}

As shown in Table~\ref{table:sample}, by fitting some of the clusters excluding H$\alpha$ contribution (only using U, B, V and I bands), their age falls above the cut off for the YMC sample and in some cases, the mass falls below the mass limit. Visual inspection of the images indicates that, though the optimal fit may be without H$\alpha$ contribution, the new lower masses do not seem to visually correspond with the cluster. For example, cluster 50574, has a new log(mass) of 2.72, however the cluster appears much more massive, and the original fit of 4.09 appears more likely. Additionally, cluster 50611 has a preferred fit including H$\alpha$, however, the image shows no H$\alpha$ emission closely surrounding the cluster, so why including H$\alpha$ is preferred is unclear, though it may be due to the continuum. 

It is clear, however, that despite improvement for many clusters by tailoring the fit to their morphology, a problem still exists with the degeneracy within the colour-colour plot and fitting procedures for ages of the clusters. The youngest parts of the model track at $\sim2$ Myr progress very rapidly. This means that, according to the model, clusters are unlikely to be found in this region, and these ages are less likely to be assigned to clusters. Additionally the degeneracy caused by clusters moving through several sections of the model track when corrected for extinction mean that ages can easily be assigned incorrectly. 

In order to investigate this effect, ages for all the clusters in the original catalogue were obtained using SED fitting both including and excluding H$\alpha$ flux. The young exposed and partially embedded clusters were selected and the $\chi^2$ values for the fit were inspected. The fit that produced the minimum $\chi_\nu$$^2$ was selected as the first optimal fit for the cluster. This way of selecting an appropriate fit proved to be inaccurate for the majority of the clusters as most of the embedded clusters were also assigned non-H$\alpha$ fits, clearly in contradiction with their morphology. These fits also put the embedded clusters at a much more advanced age than the exposed clusters, which is unphysical. Having found that this method was not entirely reliable, a more robust selection process was carried out for the fits. A youth indicator that is linked directly to the stars is therefore preferred over line emission. A potential indicator is the UV band, which will be explored further by \cite{cal14} and the HST Legacy Extragalactic UV Survey (LEGUS). 

The position of the cluster in the colour-colour plot, its SED and extinction were all used in determining the optimum fit for the cluster. Visual inspection was also used to confirm whether the fit seemed reasonable considering the H$\alpha$ morphology and the visually apparent mass of the cluster (higher mass clusters should be noticeably brighter and larger in radius, though this is a  rough technique used only to help confirm a fit). For example, an exposed cluster may be assigned a non-H$\alpha$ fit due to the $\chi^2$ value, however this fit does not account for nearby H$\alpha$ emission (if the cluster is newly exposed and gas is still close), which can be evident from the strength of the H$\alpha$ in the cluster's SED. A very large mass may be assigned to the cluster using this fit, when the cluster is clearly a lower mass cluster from only visual inspection. The SEDs proved the most valuable tool in this fit and could even possibly identify the progress of a partially embedded cluster. Later stage partials were identified from having very similar SEDs to exposed clusters (much brighter in blue bands) but with slightly higher H$\alpha$ due to nearer gas. Earlier partials looked more akin to embedded clusters with high H$\alpha$ and faint bluer bands, but could be distinguished by lower extinction. Several clusters were identified as partially embedded, despite appearing embedded, due to a lack of strong H$\alpha$ and bright blue bands. These seemed to be the clusters that had cleared gas along our line of sight but not necessarily in all directions. By employing all of these methods, a new optimum fit was assigned to each cluster, as shown in Table~\ref{table:sample}.

Partially embedded clusters are the most difficult to fit. It is often unclear whether a H$\alpha$ or non-H$\alpha$ fit is most appropriate as they can vary in H$\alpha$ morphology. This means that many of the ages for the partially embedded clusters are unreliable and have been affected by the degeneracy in the models due to extinction.

Embedded clusters can also be affected by degeneracies in the fitting procedures. Our results show that clusters are initially exposed by at least 6 Myr (and most likely much younger), and so it is unrealistic for embedded clusters, at an earlier evolutionary phase, to be older than exposed clusters, as a few are in our initial fitting. Due to their high extinction, they are prone to be assigned older ages due to the degeneracy in the model track, meaning there are several different potential ages for the clusters. In these cases, even the H$\alpha$ fits cannot obtain the younger ages that should likely be assigned to these clusters, therefore the ages provided for a small sample of the embedded clusters in the sample may be too large.

Using methods like those described previously; considering SEDs, visual inspection of the image and position on the colour-colour plot to determine which fit is appropriate, provide a more accurate age estimate for the cluster than only SED fitting. We identified 17 exposed clusters that were better fitted without H$\alpha$ and were reassigned ages. Most of these ages were only marginally different from the cluster's previous age, with only 7 clusters experiencing a fairly significant change in age. Several of the ages were older, in contrast with the apparent age of the cluster from the H$\alpha$ morphology, though the majority were given younger ages. In addition to the exposed clusters, three partially embedded clusters were also assigned different ages. 

Fig.~\ref{fig:newages} shows the new age distribution after selecting the correct fits for the clusters. The new distributions peak a lot closer together than in the previous plot and there is overlap between the different populations. The distribution for the embedded population of clusters has not changed, as they were consistently fitted with H$\alpha$. Their ages would not have improved without H$\alpha$, as they would have been assigned lower extinction and older ages due to the model degeneracy. The partially embedded clusters now have two peaks, including a younger sub-population. The latter peak consists of those with an optimal fit including H$\alpha$, while the new younger peak are those without. The exposed clusters now also have several peaks. The oldest dominant peak are those still fitted with H$\alpha$. The two younger peaks align with the embedded cluster distribution, indicating a diverse range of interaction between the stellar component of the cluster with its gas. It could potentially indicate that some clusters become exposed at a faster rate than others. The overlap between different categories of clusters could also be another indicator that the transition from embedded to exposed phase is very short, and confirms that the partially embedded clusters represent an intermediate phase of cluster evolution to an exposed state. It also shows that this process is continuous and that a majority of the clusters have a compact bubble of ionised gas by $\sim3$ Myr. 

\begin{table*}
Final sample of young massive clusters in M83
\centering
\begin{tabular}{c c c c c c c c c c}
\hline
Cluster ID & Cat & SED log(age) & Log(mass) & $A_{V}$ & Max distance & Min distance & \% max & \% min & Best fit\\
& & (Best fit) & (Best fit) & & (pc) & (pc) & & & \\
\hline\hline
10141$^\ast$	&	1	&	6.60	&	4.03	&	0	&	n/a	&	n/a	&	n/a	&	n/a	& No H$\alpha$	\\
10165	&	2	&	6.70	&	4.00	&	0	&	$>80$	&	23.1	&	46	&	10	& No H$\alpha$	\\
10426	&	0	&	6.60	&	4.00	&	2.7	&	n/a	&	n/a	&	n/a	&	n/a	& H$\alpha$	\\
10433	&	0	&	6.60	&	4.58	&	2.86	&	n/a	&	n/a	&	n/a	&	n/a	& H$\alpha$	\\
10438	&	1	&	6.85	&	4.79	&	1.52	&	28.0	&	7.8	&	4	&	18 & H$\alpha$	\\
10452	&	0 	&	6.70	&	4.18	&	3.53	&	n/a	&	n/a	&	n/a	&	n/a	& H$\alpha$	\\
10594	&	2	&	6.30	&	4.46	&	0	&	48.9	&	14.9	&	9	&	19 & No H$\alpha$	\\
10597	&	2	&	6.60	&	4.72	&	0	&	54.9	&	10.0	&	5	&	22 & No H$\alpha$	\\
10645	&	2	&	6.60	&	3.97	&	0	&	$>80$	&	10.0	&	28	&	31 & No H$\alpha$	\\
10732	&	2	&	6.60	&	4.03	&	0	&	$>80$	&	20.0	&	21	&	28 & No H$\alpha$	\\
10915	&	1	&	6.75	&	3.73	&	2.78	&	10.0	&	0	&	16	&	83 & H$\alpha$ \\
10957	&	0	&	6.73	&	3.57	&	2.08	&	10.0	&	0	&	11	&	89 & H$\alpha$	\\
11164	&	2	&	6.60	&	4.01	&	0	&	$>80$	&	26.5	&	16	&	28 & No H$\alpha$	\\
11174$^\ast$	&	1	&	6.73	&	3.95	&	0.14	&	n/a	&	n/a	&	n/a	&	n/a	& H$\alpha$\\
11266	&	1	&	6.88	&	3.82	&	0.83	&	30.9	&	0	&	17	&	21	& H$\alpha$ \\
11297	&	1	&	6.60	&	4.06	&	0	&	$>80$	&	0	&	14	&	13 & No H$\alpha$	\\
11305	&	1	&	6.70	&	3.81	&	0.46	&	13.8	&	0	&	8	&	92 & H$\alpha$	\\
11312	&	0	&	6.60	&	3.92	&	2.7	&	n/a	&	n/a	&	n/a	&	n/a	& H$\alpha$\\
11414	&	2	&	6.70	&	3.87	&	0	&	$>80$	&	67.2	&	90	&	10	& No H$\alpha$\\
20249	&	0	&	6.48	&	3.98	&	3.54	&	n/a	&	n/a	&	n/a	&	n/a	& H$\alpha$	\\
20281	&	2	&	6.60	&	4.08	&	0	&	23.1	&	10.0	&	8	&	17	& No H$\alpha$	\\
20288	&	2	&	6.78	&	4.08	&	0	&	$>80$	&	40.0	&	45	&	14 & No H$\alpha$	\\
20314	&	1	&	6.73	&	3.81	&	0	&	30.2	&	0	&	5	&	15	& No H$\alpha$\\
20417	&	2	&	6.60	&	4.26	&	0	&	$>80$	&	40.0	&	20	&	38	& No H$\alpha$\\
20866	&	2	&	6.88	&	3.75	&	0.02	&	$>80$	&	15.6	&	33	&	35 & H$\alpha$	\\
20868	&	2	&	6.88	&	3.80	&	0.18	&	$>80$	&	14.8	&	33	&	35 & H$\alpha$	\\
20953	&	1	&	6.73	&	3.70	&	0.75	&	$>80$	&	53.2	&	63	&	13 & H$\alpha$	\\
30126	&	1	&	6.87	&	3.97	&	0.79	&	23.4	&	0	&	18	&	40	& H$\alpha$\\
30701	&	0	&	6.60	&	3.82	&	2.19	&	n/a	&	n/a	&	n/a	&	n/a	& H$\alpha$\\
30989	&	0	&	6.30	&	3.98	&	2.29	&	n/a	&	n/a	&	n/a	&	n/a	& H$\alpha$\\
31015	&	2	&	6.60	&	3.93	&	0	&	$>80$	&	15.0	&	13	&	18	& No H$\alpha$\\
31016	&	2	&	6.70	&	4.00	&	0	&	$>80$	&	20.0	&	9	&	5 & No H$\alpha$	\\
40027	&	0	&	6.48	&	4.19	&	3.77	&	n/a	&	n/a	&	n/a	&	n/a & H$\alpha$	\\
40117	&	2	&	6.87	&	4.34	&	0.54	&	$>80$	&	15.2	&	6	&	30 & H$\alpha$	\\
40358	&	2	&	6.88	&	3.89	&	1.22	&	$>80$	&	24.0	&	32	&	7 & H$\alpha$	\\
40543	&	2	&	7.78	&	3.37	&	0	&	$>80$	&	46.4	&	90	&	7 & No H$\alpha$	\\
40820	&	2	&	6.95	&	4.20	&	0.38	&	$>80$	&	56.8	&	94	&	6 & H$\alpha$	\\
40840	&	2	&	7.30	&	3.23	&	0	&	$>80$	&	44.9	&	78	&	7 & No H$\alpha$	\\
40915	&	2	&	6.78	&	3.70	&	0	&	23.7	&	10.0	&	27	&	11	& H$\alpha$\\
40926	&	2	&	6.78	&	4.1	&	0	&	37.5	&	7.9	&	16	&	41	& H$\alpha$\\
50015	&	2	&	6.78	&	3.88	&	0	&	40.0	&	5.5	&	3	&	20 & H$\alpha$	\\
50019	&	2	&	6.79	&	3.74	&	0	&	$>80$	&	13.3	&	45	&	7 & H$\alpha$	\\
50035	&	2	&	6.60	&	2.50	&	0.78	&	$>80$	&	10.0	&	12	&	43	& No H$\alpha$\\
50081	&	1	&	6.88	&	3.77	&	0.39	&	35.3	&	0	&	14	&	52	& H$\alpha$\\
50099	&	0	&	6.60	&	3.73	&	1.99	&	n/a	&	n/a	&	n/a	&	n/a	& H$\alpha$	\\
50140	&	2	&	6.30	&	4.21	&	0	&	$>80$	&	43.1	&	60	&	11 & No H$\alpha$	\\
50160	&	1	&	6.70	&	3.70	&	0	&	n/a	&	n/a	&	n/a	&	n/a	& H$\alpha$	\\
50233	&	2	&	6.78	&	4.04	&	0	&	$>80$	&	7.6	&	21	&	28 & H$\alpha$	\\
50256	&	2	&	6.87	&	3.81	&	0	&	$>80$	&	28.4	&	24	&	10	& H$\alpha$\\
50358	&	0	&	6.70	&	3.96	&	3.11	&	n/a	&	n/a	&	n/a	&	n/a	& H$\alpha$\\ 
50377	&	1	&	6.78	&	3.94	&	0	&	$>80$	&	0	&	11	&	22 & H$\alpha$	\\
50574	&	2	&	6.78	&	4.09	&	0	&	$>80$	&	45.4	&	79	&	3 & H$\alpha$	\\
50611	&	2	&	6.78	&	3.70	&	0	&	$>80$	&	14.8	&	5	&	13	& H$\alpha$\\
50789	&	1	&	6.60	&	4.73	&	0	&	31.5	&	0	&	4	&	41 & No H$\alpha$	\\
50905	&	0	&	6.60	&	3.70	&	0.33	&	n/a	&	n/a	&	n/a	&	n/a	& H$\alpha$\\
51080	&	2	&	6.78	&	4.11	&	0	&	48.3	&	10.0	&	3	&	31 & H$\alpha$	\\
51093	&	2	&	6.78	&	3.84	&	0	&	49.2	&	7.4	&	2	&	22	& H$\alpha$\\
51095	&	2	&	6.78	&	3.96	&	0	&	40.0	&	10.0	&	16	&	5 & H$\alpha$	\\
60171	&	0	&	6.48	&	3.73	&	2.15	&	n/a	&	n/a	&	n/a	&	n/a	& H$\alpha$\\
60558	&	2	&	6.78	&	3.90	&	0	&	$>80$	&	8.3	&	20	&	23	& H$\alpha$\\
60571	&	2	&	6.80	&	4.28	&	0	&	$>80$	&	7.6	&	8	&	31 & H$\alpha$	\\
\hline
\pagebreak

\end{tabular}
\end{table*}

\begin{table*}
\begin{tabular}{c c c c c c c c c c}
\hline
Cluster ID & Cat & SED log(age) & Log(mass) & $A_{V}$ & Max distance & Min distance & \% max & \% min & Best fit \\
& & & & & (pc) & (pc) & & &\\
\hline\hline
60582	&	0	&	6.60	&	4.06	&	3.48	&	n/a	&	n/a	&	n/a	&	n/a	& H$\alpha$\\
60770	&	1	&	6.79	&	4.09	&	0	&	73.9	&	0	&	10	&	24 & H$\alpha$	\\
61053	&	0	&	6.70	&	3.87	&	3.08	&	n/a	&	n/a	&	n/a	&	n/a	& H$\alpha$\\
61216	&	2	&	7.00	&	4.13	&	0.07	&	$>80$	&	16.1	&	21	&	24	& H$\alpha$\\
61310	&	1	&	6.92	&	3.81	&	0.32	&	$>80$	&	4.5	&	5	&	11	& H$\alpha$\\
61417	&	2	&	6.78	&	4.57	&	0	&	$>80$	&	12.5	&	10	&	25 & H$\alpha$	\\

\hline
\end{tabular}
\caption{Properties of the final clusters in the YMC sample of M83, after removing poor sources. Cluster ID corresponds to the ID in the main Silva-Villa et al, 2014 catalogue, with the first digit indicating the field where the cluster is located. The 'cat' indicates the category the cluster was sorted into: '0', '1' and '2' correspond to embedded, partially embedded and exposed respectively. Clusters indicated with a $^\ast$ are clusters that initially appear to be embedded from the H$\alpha$ morphology, though are partially embedded when you consider their SEDs, position on the colour-colour plot and their extinction. They seem to have cleared gas within our line of sight so they appear more exposed without visual inspection of the image. 'Max distance' is the furthest radial distance cleared of gas by the cluster, 'Min distance' is the minimum. '\% max' and '\% min' relate to the percentage of the circular region surrounding each cluster cleared to the maximum and minimum distances. Measurements of these properties were not obtained for embedded clusters. 'Best fit' is the optimum SED fit for the cluster after checking each cluster's individual SED, position on the colour-colour plot and H$\alpha$ morphology. 'H$\alpha$' is a fit with U, B, V, H$\alpha$ and I bands and 'No H$\alpha$' includes U, B, V and I.}
\label{table:sample}
\end{table*}

\subsubsection{Infrared age dating of clusters}
\label{sec:hband}

Space-based resolution capabilities will soon be limited to near-IR data and the traditional extragalactic cluster population studies in optical and UV wavelengths will no longer be feasible with the retirement of HST. These clusters should be visible in the near-IR, however studies that attempted to use infrared data, e.g \cite{gd13}, using aperture photometry as with optical data, have proven unreliable due to spatial resolution effects \citep{nb14}. Despite this, space-based instruments with high spatial resolution (e.g. JWST) in the near-IR can still be relevant and useful to cluster studies.

As explained previously, current age dating methods for clusters, such as SED fitting, as used for this sample is poor at providing accurate ages for clusters younger than $\sim10$ Myr. Such inaccuracy means that any errors associated with studies of YMCs with ages in this range are likely to be significant. 

A novel method for providing a more accurate age, or for confirming previous age estimates has been developed by \cite{gaz11}, which uses the presence of red supergiants (RSGs) in the cluster as an absolute age indicator. By plotting the J-H colour for each of the clusters against log(age) (obtained via SED fitting), a clear distinction between younger and older clusters exists in colour space. Younger clusters that are devoid of RSGs appear distinctly bluer, with values jumping to redder colours at older ages, immediately at the onset of RSGs. This is due to RSGs dominating the flux at red wavelengths, such as the near-IR. According to \cite{gaz11}, any clusters with red J-H colours should be older than $\sim6$ Myr, the age of the jump in the plot. One caveat for this method is that extinction can also cause young clusters to have red J-H colours.  

\begin{figure}
\includegraphics[width=8.5cm]{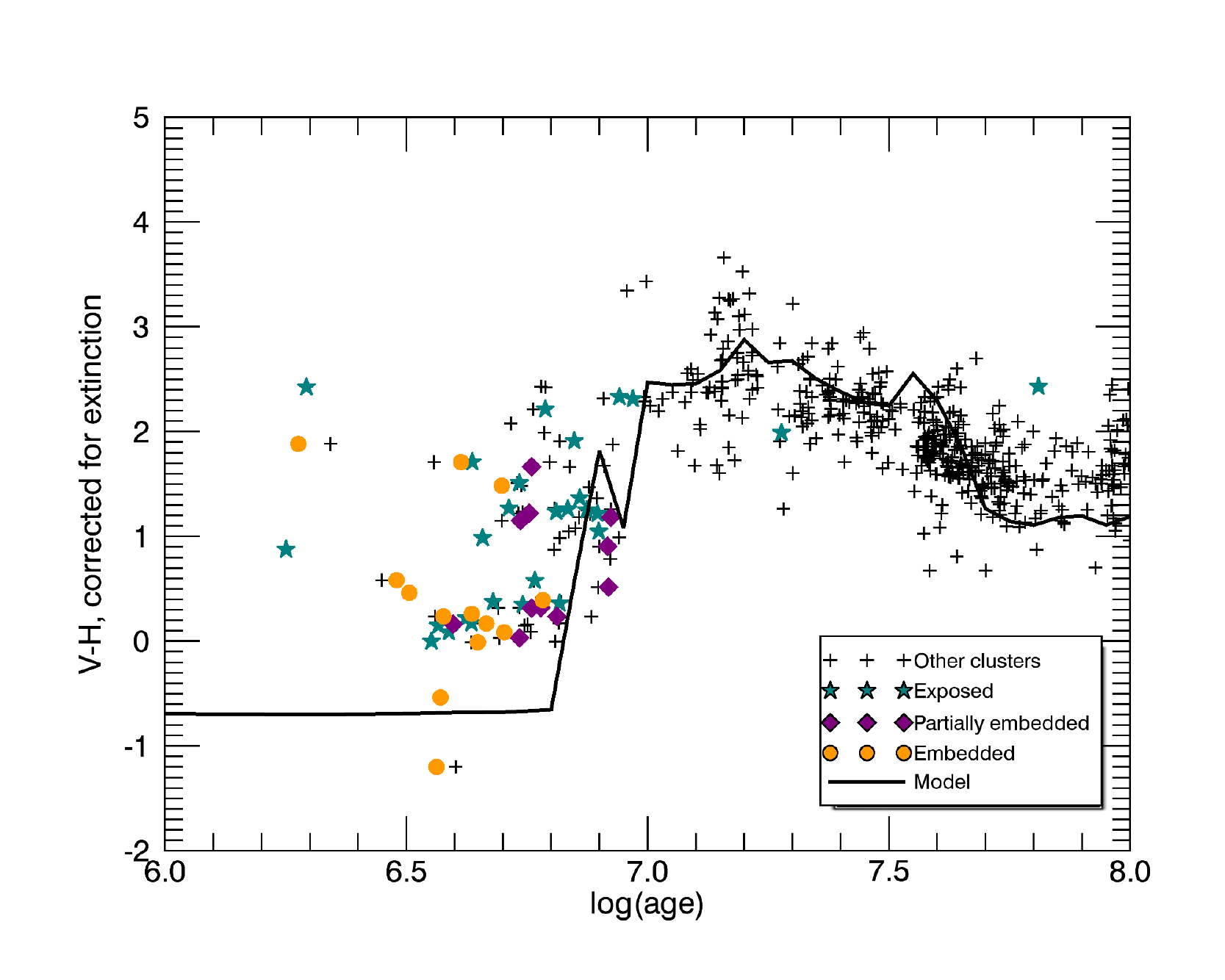}
\caption{V-H colour for all clusters in the M83 cluster sample, with the subsample of YMCs highlighted in colour with green stars, purple diamonds and orange circles corresponding to exposed clusters, partially exposed clusters and embedded clusters respectively. The jump from bluer, younger clusters to redder, older clusters due to the appearance of red supergiants is clear, with the exposed clusters at redder colours due to their slightly older age, and the youngest embedded clusters as the bluest. The model plotted over the data is calculated for the V-H band as per \protect\cite{gaz11}.} 
\label{fig:vhplot}
\end{figure}  

For this study, the log(age) (based on our best fit, as quoted in Table~\ref{table:sample}) was plotted against V-H colour (which was corrected for extinction) to find whether this jump is visible. The V band was chosen as no J band data was available for WFC3 in Fields 3, 4, 5, 6 and 7. The plot is shown in Fig.~\ref{fig:vhplot}, and despite the difference in wavelength, the same feature is produced with the jump in colour at the same age, $\sim6$ Myr. Almost all of the YMCs are where expected; older exposed clusters are redder in V-H, as they have RSGs and embedded younger clusters are the bluest. This indicated that our age estimates are fairly accurate, and at the values we would expect for these clusters. Though our SED fitted ages are rough estimates, this trend is still clearly observed, indicating the robustness of this technique. 

\subsubsection{Wolf-Rayet features in clusters}
\label{sec:wolfrayet}

Wolf-Rayet (WR) stars are the later evolutionary stages of massive stars (typically 10-25 \msun) and can be detected in clusters by broad emission lines in their spectra that are specific to these types of stars. The majority of WR stars are derived from O-type stars and so have total lifetimes of $\sim$ 5 Myr, 10 \% of which is spent in a WR phase. Higher mass (25-30 \msun) RSGs can also become WR stars. The characteristic emission lines arise from fast stellar winds and heavy mass loss from the star \citep{crowther07}.

The presence of such high mass, short-lived stars in clusters within our sample indicates that the clusters should be young. In older clusters, any WR stars would have already evolved past supernovae, given their $\sim$ 5 Myr lifetimes, giving us a refined age estimate of these clusters between 4-7 Myr. \cite{ba06} previously used the same technique to age date clusters in the Antennae galaxies.

\cite{hadfield05} carried out a survey searching for WR stars in regions within M83, discovering 132 sources. We cross-referenced the coordinates of the regions they identified as containing these objects with the coordinates of our cluster sample, examining images of the position of the features to confirm that the features originated from the cluster. We found 5 regions that coincided with the position of clusters in the YMC sample. In addition to these 5 clusters, we also had optical spectroscopy of one further cluster in the sample that showed WR features. These clusters and their properties are shown in Table~\ref{table:wr}. 

\begin{table}
	\centering
	Properties of clusters with WR features
	\begin{tabular}{c c c c}
		\hline
		Cluster ID & Cat & Log(Age) & Log(Mass) \\
		\hline \hline
		10141 & 1 & 6.60 & 4.03 \\
		10438 & 1 & 6.85 & 4.79 \\
		10597 & 2 & 6.60 & 4.72 \\
		31015 & 2 & 6.60 & 3.93 \\
		50160 & 1 & 6.70 & 3.70 \\
		51080 & 2 & 6.78 & 4.11 \\
		\hline
	\end{tabular}
	\caption{Clusters that coincide with identified WR regions in M83. 'Cat' refers to classification of the cluster as embedded, partially embedded or exposed, as used throughout the paper. The ages and masses are the revised figures as described in \S~\ref{sec:nohafit}.}
	\label{table:wr}
\end{table}

The low number of sources corresponding to clusters in our sample is mostly due to the lack of coverage of the original study. Their spectra did not uniformly cover the entire galaxy and so many clusters will have been missed. Additionally, these stars will be harder to detect in clusters due to being "washed out" by the continuum of the cluster, so many WR sources will be required for a detection. There are many young massive clusters that do not align with the WR sources, although it is not guaranteed that all clusters should contain enough WR stars to make detection possible. So the clusters without corresponding WR features may not be older, they just may not have sufficient WR stars to be detected. Therefore the WR feature method can only be used to confirm young ages for some clusters but not decide on the correct age. We believe the other methods discussed in this section should primarily be used for this purpose.

Using this method indicates that 5 of the clusters with WR features have been correctly assigned younger ages. It also suggests that cluster 10438 should be younger. This cluster is already thought to have been given too old an age due to degeneracy in the model track, and this test agrees with that conclusion. The sample of 6 clusters used here show that WR features can be used to constrain ages of young clusters.

\section{Results}
\label{sec:results}

\subsection{Statistical age results}

After comparison of the categorisations of the clusters with their age estimates, the results indicate that clusters are no longer embedded, and have removed their gas by $\sim4$ Myr. The median ages for the embedded, partially embedded and exposed clusters were $\sim4$ Myr, $\sim5$ Myr and $\sim6$ Myr, respectively. Therefore, the age at which clusters become exposed should be between 4 and 5 Myr. The maximum age of the clusters in the embedded sample is $\sim6$ Myr, possibly indicating that the timescale for the clusters to progress from an embedded to exposed state is very short. Alternatively, this could just be an indication of the inaccuracies of SED fitting at ages $<$ 10 Myr, as discussed in \S~\ref{sec:nohafit}. Older ages for the embedded and partially embedded clusters could arise from model degeneracies in SED fitting. These incorrect ages may have skewed the median ages for embedded and partially embedded clusters, which should in fact be younger. This issue can be somewhat mitigated by determining the correct fit, taking into account that embedded clusters should be younger than exposed clusters (therefore younger than $\sim6$ Myr), as per the results of this paper.

The majority of the clusters (53\%) in the sample are exposed. If we include clusters that have removed any amount of gas from their surroundings i.e. partially embedded clusters as well, this means that 77\% of the sample are no longer embedded. If we assume this sample is complete for M83 for clusters younger than 10 Myr (a 10 Myr age cut was applied to the catalogue), and that the cluster formation rate is constant, this indicates that 7.7 Myr of the first 10 Myr of the life of young massive clusters are spent exposed to some extent. Alternatively, the first 2.3 Myr are spent in an embedded state, and clusters initially become exposed at this age, in agreement with the age indicated by the colour-colour plot discussed in \S~\ref{sec:ccplot}.

A key assumption of the calculation, however is that all clusters are optically visible and enter our catalogue when they form. It can be argued that far younger clusters may be completely obscured and not be included, however a study by \cite{whit02} of clusters in the Antennae indicates that by comparing compact radio sources (extremely young clusters) with optical clusters, there is $\sim85\%$ overlap. This indicates that very few of these sources would be missed. Additionally, we searched the H band images, (as well as ground-based JHK band images in \cite{nb14}) and found no clear massive cluster progenitors that were not already included in our optical sample. This is also an indication that the embedded phase of the clusters is very short and that the clusters are very efficient at removing gas.

\subsection{Models of bubble expansion}
\label{sec:model}

\cite{dale14} carried out simulations of cluster formation and early evolution. Their simulations include the effects of both ionisation and stellar winds from OB stars, although no radiation pressure has been included, which is likely to be relevant for very dense star-forming regions, e.g. \cite{fall10}. The simulations end 3 Myr after the formation of the first O-stars in each model cloud and some simulations already display bubbles formed by clusters with  masses similar to the lower end of the range represented by our sample. This agrees with our results that these clusters will have removed their gas by 1-3 Myr. However, the bubbles created in \cite{dale14} are substantially smaller than those we observe.

Having measured the maximum distance cleared by each of the exposed clusters, 77\% of clusters had cleared some extent of their gas to radii greater than 80 pc. In contrast, simulations of clusters with similar initial masses clear bubbles to a maximum possible distance of 40-50 pc in diameter. This indicates that the simulations do not reproduce the extremely efficient process these clusters use to remove gas. It could mean that the process is related to radiation pressure, as this was not included in the simulations. Only a few of the simulations in \cite{dale14} are likely to be dense enough for radiation pressure to be important, but that may not be true of the clusters studied here. Alternatively, it could be an already known physical process that is far more effective than previously thought (such as stellar winds), or a completely new process.

An important note is that these simulations end 3 Myr after the birth of the first O-stars. Should they run until 6 Myr, the size of the bubbles in the simulations will increase and likely agree with our measurements. Therefore, we conclude that some of the simulations of gas removal are in reasonable agreement with what we find in the young clusters in M83.

\subsection{Cluster mass, distance cleared and cluster blowouts}
\label{sec:blowouts}
\begin{figure}
\includegraphics[width=8.5cm]{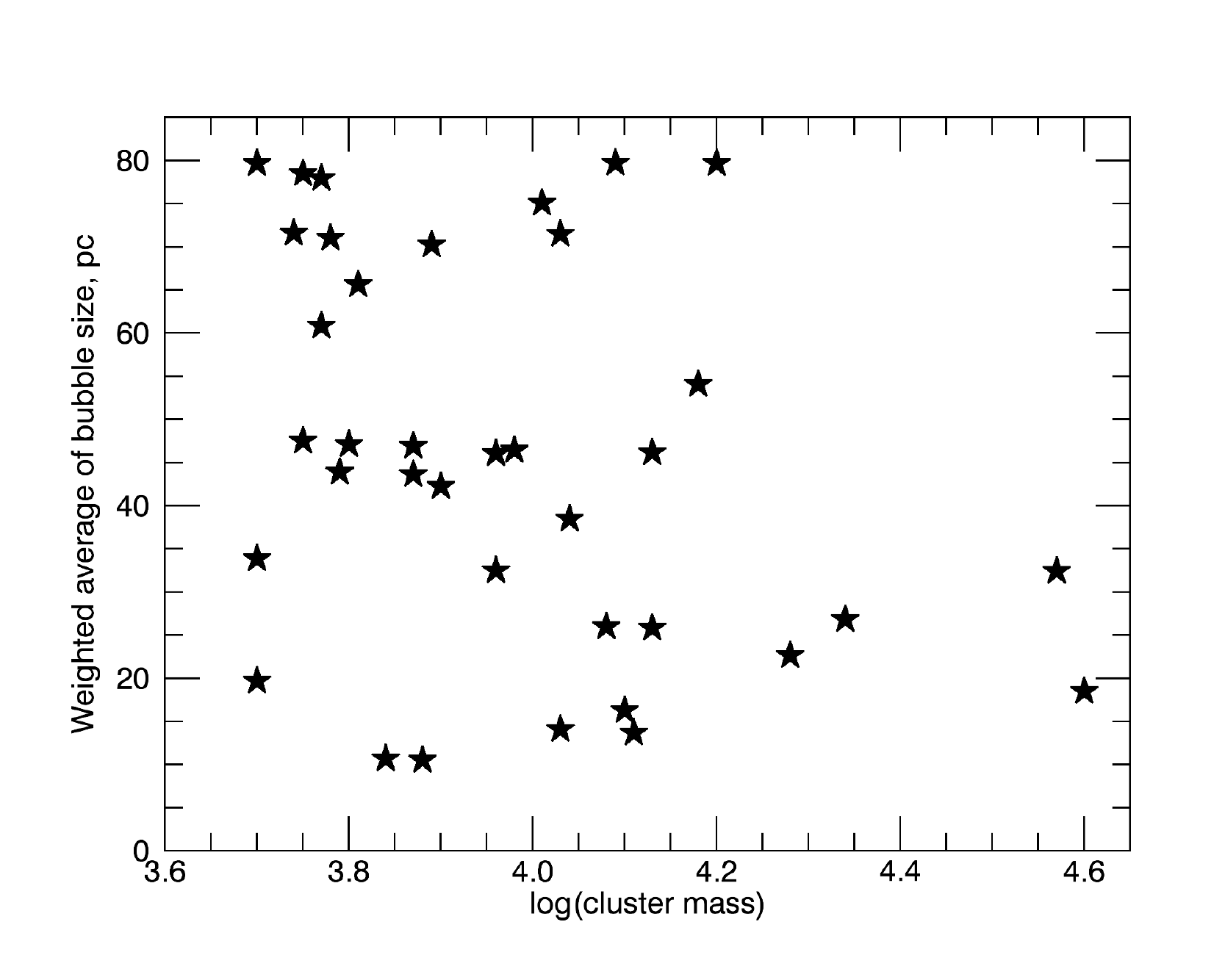}
\caption{The plot of log mass of the cluster against the weighted average of the interior size of the bubble for exposed clusters. As shown, there appears to be no correlation between the two variables, and may require a third variable to constrain a relationship.}
\label{fig:massdistance}
\end{figure}

An interesting aspect of this study to take further, is whether or not there is a relationship between the mass of the cluster and the size of the bubble cleared. Using the measurements of the amount of space cleared by the cluster from Table~\ref{table:sample}, we looked at the plot of mass of the cluster against the weighted average of the maximum and minimum distances cleared, as shown in Fig.~\ref{fig:massdistance}. There was no clear correlation between the amount cleared and the mass of the cluster, for clusters in our mass range of $\sim5000 - 6x10^4$ \msun. It is probable that if a relationship exists, it also requires a third variable to constrain it, such as position of the cluster in the galaxy (whether upstream or downstream of an arm, or in the inter-arm regions), or the cluster age. Including the cluster's age in our plots still did not constrain any relationship, so we then decided to test the inclusion of galactic position.  

To test this theory, we visually identified the position of each cluster in the sample. This property was quantised into 6 possible categories: in the bulge/inner bar region, in the outer bar region, at the end of the bar (which has increased levels of star formation), in the upstream regions of the arm just after gas enters the arm region and is shocked (leading to star formation), in downstream regions at the arm's outer edge, after the expected star formation, or outside of the arm overdensities. We plotted cluster mass against the weighted average distance of the cleared bubble as a function of position in the galaxy and still no clear relationship could be determined. 

Including the mass of the gas outside of the exposed clusters could help constrain a possible relationship. However, obtaining an accurate estimate of the amount of gas associated with the cluster would be very difficult using images that rely on photometric bands highly susceptible to the continuum. Furthermore, the physical size of the shell would be very hard to determine, especially in more crowded environments. 

An additional aspect of the clusters that can be studied is the presence of a `blowout' in the gas morphology. We would define a blowout as free-streaming in one direction within the gas cleared. Each of the exposed and partially embedded clusters were visually inspected to obtain an estimate of the number of clusters within the sample that contain a blowout region. We found that 18\% of the entire cluster population had a noticeable blowout in the gas. As it would be impossible for embedded clusters to have a blowout (they have not yet strongly affected their environment), when considering only partially embedded and exposed clusters, 24\% of the population had a blowout. This indicates that this phenomenon should be fairly common for clusters of this mass. \cite{pell12} also studied blowouts in high mass (luminosity) HII regions in the LMC and found all regions were leaking photons.  

\subsection{Implied constraints on age spreads within clusters}
\label{sec:agespreads}

Phenomena observed in intermediate age (1-2 Gyr) clusters in the LMC and SMC, such as double, or extended main sequence turn offs, e.g. \cite{mack07}, \cite{goud09}, \cite{girardi13}, and in globular clusters such as split main sequences, e.g. \cite{piotto05}, \cite{milone13}, and chemical spreads have widely been explained using age spreads, or multiple bursts of star formation. The extent of the age spreads can be up to a few hundred Myr (e.g. \cite{milone09}), though this has been strongly contested by \cite{nb13} and \cite{nieder15}, who have found no evidence of age spreads in YMCs with resolved photometry. Additionally, stellar rotation has been shown to potentially exhibit the same effects seen in the main sequence turn offs \citep{nb09}. \cite{ivan14} also showed that the spectra of NGC 34 can only be reliably fit with a single stellar population, contesting the theory of multiple bursts of star formation. 

The debate of the origin of multiple stellar populations continues, and our results can be used to apply some constraints to current claims of age spreads. The clusters studied here are far younger than those with the proposed age spreads, and they have already become exposed and free of gas at 4 Myr. The lack of ionised gas within the cluster at this time means that there can be no star formation until more gas has been accreted, although the mechanism used to accrete this gas is still unclear.

As shown in \cite{ba14}, this is also true of very young clusters, of approximately the same age but larger masses than the clusters studied in this paper. In that study it was shown that the gas was still moving away from the clusters at a speed of $\sim$ 40 km/s, and therefore the pristine gas required to produce the second generation of stars would need to be accreted from other sources, unrealistic for the amount required. Our clusters show very similar results. The removed gas should be receding from the cluster at a speed of $\sim$ 20 km/s, and clusters of this mass would not be able to efficiently re-accrete this gas in the proposed time between the two generations forming. It is highly unlikely the cluster would be able to accrete this required material from elsewhere in the galaxy on this timescale (e.g. \cite{nb14}) also, especially considering the amount of gas required would be on par with the existing cluster mass for a reasonable star formation efficiency. This indicates that the current required age spreads within clusters are unrealistic for young massive clusters in this sample.

\section{Discussion}
\label{sec:discussion}

The results of this paper have important consequences for one of the current globular cluster formation scenarios, the Fast Rotating Massive Star Scenario (FRMS) \citep{krause13}. In this scenario, the first generation of stars form, which then become mass segregated leaving the high mass stars in the centre. Bubbles of gas surround each high mass star in narrow filaments. Decretion discs form around the high mass stars and ejected material mixes with the pristine gas from the filaments to form the second generation of stars within the disks. When the high mass stars have evolved to the supernova stage, the leftover pristine gas from the filamentary structures is arranged in a larger bubble surrounding all stars in the cluster, so that the cluster appears to be embedded in gas.    

The young clusters must remain embedded in their natal gas for 25-30 Myr in the FRMS scenario, however the clusters in this study have shown that this is not the case, and they have in fact removed their excess gas by $\lesssim4$ Myr. This clearly indicates that the FRMS scenario, in its current form, is not feasible and clusters are far more efficient at removing gas than expected. One clear difference is the mass of the cluster that has formed. The FRMS scenario discusses clusters of mass $\gtrsim10^6$~\msun, whereas the most massive cluster in this work is $\sim10^4$~\msun. However, the results of \cite{ba14} suggest that the effect of clusters clearing out gas is scalable between cluster masses. They present similar results for clusters of masses $\sim10^6$~\msun, on par with those described in the FRMS scenario. The same effect is seen in these clusters; gas is cleared at the same age as for the lower mass clusters. One difference is the size of the bubble cleared. Higher mass clusters appear to create larger bubbles on the same timescales, probably due to larger feedback acting to remove the gas. 

The mechanism acting within the clusters to remove the gas is much more significant than previously assumed, shown by the clusters ability to clear gas so early in their evolution. Supernovae have been considered as one of the main contributors to the removal of gas. It is feasible to include supernovae for systems at approximately solar metallicity, like M83, however this only holds if the cluster contains very high mass stars. For the first supernovae to occur at 2-3 Myr after cluster formation (the age at which the clusters seem to start removing gas) the IMF must be well sampled. A reasonable estimate for the most massive star in clusters of $\sim5000$ \msun\ is $\sim40$ \msun, which would not undergo a supernova phase until $\sim8$ Myr, and so will not have made an impact on the removal of gas. Therefore for supernovae to be considered, stars much greater than this mass (likely at least one star of $\sim120$ \msun) must exist in each of the clusters, which is unlikely to be the case for all clusters. At sub-solar metallicities, these very high mass stars would collapse straight into a black hole phase, but supernovae could at least make a contribution to the gas removal at solar metallicity, though they are very unlikely to be the sole mechanism \cite{heger03}.

\section{Conclusions}
\label{sec:conc}

Here we summarise our main conclusions:

\begin{itemize}
\item{The young massive clusters in our sample are all exposed and free of gas at $\sim4$ Myr. The process of removing gas seems to begin at 2-3 Myr, as indicated by the position of exposed clusters on in colour-colour space and statistical considerations of the relative population sizes of exposed or partially embedded clusters as opposed to embedded clusters.}
\item{The mechanism for the removal of gas from clusters is far more efficient than previously thought. Supernovae could play a role but cannot be the sole contributors. Radiation pressure could contribute, as well as strong stellar winds from stars at advanced evolutionary stages, such as Wolf-Rayet stars, which have been identified in several of the YMCs in the sample.}
\item{Current models of gas clearance from clusters could potentially produce similarly sized bubbles when compared with our images, though radiation pressure was not included and could be a key contributor to the removal of gas.}
\item{The lack of gas within these clusters provides constraints for the age spreads that have been proposed to exist. A spread of a few hundred Myr is required, however these clusters have removed their gas by $\sim4$ Myr, and so cannot form a new generation of stars, unless this gas is re-accreted, or new pristine gas is obtained. The amount of gas that would need to be accreted over this short time frame is unlikely, especially considering the cluster's original gas is pushed away to large distances, out of reach for re-accretion. It is also unlikely that gas from other sources could be accreted in the space of a few hundred Myr. This also has consequences for the FRMS scenario which currently requires clusters to remain embedded in natal gas for 25-30 Myr in order to form a second generation of stars is infeasible, as these clusters have removed gas by $\sim4$ Myr.}
\item{Age fitting using SEDs is often unreliable for young clusters due to degeneracies in the models and aperture effects. Constraints on age can be made by using the near-infrared bands, such as H (F160w), where the characteristic jump at $\sim5-6$ Myr from blue to red colours due to RSGs can provide an upper limit. The presence of Wolf-Rayet features also puts a constraint on age of the cluster of 4-7 Myr. Using H$\alpha$ in the fit for the age of a young cluster and as an indicator of age is very unreliable as it is not directly linked to the stars, but rather line emission and the continuum can have an effect. A better indicator could be the UV.}
\item{No relationship has been found between the mass of the cluster and size of the bubble cleared for clusters of mass between 5000 and $6x10^4$ \msun}. If a relationship does exist, it likely requires more variables to constrain the correlation.
\item{Approximately 25 \% of the clusters were identified as having 'blowouts'. This indicates that it is a fairly common phenomenon in clusters.}

\end{itemize}

\section*{Acknowledgments}

Based on observations made with the NASA/ESA Hubble Space Telescope, and obtained from the Hubble Legacy Archive, which is a collaboration between the Space Telescope Science Institute (STScI/NASA), the Space Telescope European Coordinating Facility (ST-ECF/ESA) and the Canadian Astronomy Data Centre (CADC/NRC/CSA).

We thank William Blair for providing continuum subtracted H$\alpha$ images for this study.

J. E. Ryon gratefully acknowledges the support of the National Space Grant College and Fellowship Program and the Wisconsin Space Grant Consortium.

This research was supported by the DFG cluster of excellence `Origin and Structure of the Universe' (JED).

\bibliographystyle{mn2e}

\label{lastpage}
\end{document}